\documentclass[12pt]{iopart}

\usepackage{amsfonts, amssymb, mathrsfs, times, float, color, bm}
\usepackage{graphicx}
\usepackage{graphicx,epstopdf}
\usepackage{dcolumn}
\usepackage{bm}
\newcommand{\nn}{\nonumber}

\begin{document}

\title[Light and Massive Particle Deflection by a Moving KN Black Hole]{Gravitational Deflection of Light and Massive Particle by a Moving Kerr-Newman Black Hole}

\author{Guansheng He \& Wenbin Lin}

\address{School of Physical Science and Technology, Southwest Jiaotong University, Chengdu 610031, China}
\ead{wl@swjtu.edu.cn}
\vspace{10pt}

\begin{abstract}
The gravitational deflection of test particles including light, due to a radially moving Kerr-Newman black hole with an arbitrary constant velocity being perpendicular
to its angular momentum, is investigated. In harmonic coordinates, we derive the second post-Minkowskian equations of motion for test particles, and solve them by high-accuracy numerical calculations. We then concentrate on discussing the kinematical corrections caused by the motion of the gravitational source to the second-order deflection. The analytical formula of light deflection angle up to second order by the moving lens is obtained. For a massive particle moving with a relativistic velocity, there are two different analytical results for Schwarzschild deflection angle up to second order reported in the previous works, i.e., $\alpha(w)=2\left(1+\frac{1}{w^2}\right)\frac{M}{b}+3\pi\left(\frac{1}{4}+\frac{1}{w^2}\right)\frac{M^2}{b^2}$ and $\alpha(w)=2\left(1+\frac{1}{w^2}\right)\frac{M}{b}+\left[3\pi\left(\frac{1}{4}+\frac{1}{w^2}\right)+2\left(1-\frac{1}{w^4}\right)\right]\frac{M^2}{b^2}$, where $M,$ $b,$ and $w$
are the mass of the lens, impact parameter, and the particle's initial velocity, respectively. Our numerical result is in perfect agreement with the former. Furthermore, the analytical formula for massive particle deflection up to second order in the Kerr geometry is achieved. Finally, the possibilities of detecting the motion effects on the second-order deflection are also analyzed.
\end{abstract}


\vspace{2pc}

\section{Introduction}
The time-dependence of a background metric, caused by the motion of a gravitational system, usually exerts a correctional effect on the propagation of an electromagnetic signal or celestial body. This kind of kinematical effect, which is called ``velocity effect" or ``motion effect", has been investigated in detail. In contrast to most previous works
(see, Refs.~\cite{PB1993,CLPS1999}, and references therein), which were focused on the effects in the first-order velocity (\emph{FOV}) and first-order deflection (\emph{FOD}) approximations, Kopeikin and Sch$\ddot{a}$fer~\cite{KopeiSch1999} studied light propagation in the gravitational field of an arbitrarily moving N-body system analytically. Their calculations were performed in the first post-Minkowskian (\emph{1PM}) approximation (weak-field approximation without limiting to low velocity), and were generalized in Ref.~\cite{KopeiMash2002} to consider the spin-dependent gravitomagnetic effects on the propagation of light. In 2003, the solutions of Li\'{e}nard-Wiechert potential \cite{KopeiSch1999,KopeiMash2002} were confirmed in Ref.~\cite{Klioner2003} via exerting a Lorentz boosting on the propagation equations of light in the field of a static source. In order to extend and apply the analytical results achieved in Ref.~\cite{Klioner2003} to astrometric observations, Klioner and Peip~\cite{KlioPeip2003} performed high-resolution numerical simulations for the trajectory of light in the field of moving point masses, in the first post-Newtonian approximation as well as \emph{1PM} approximation. In 2004, Wucknitz and Sperhake~\cite{WuckSperh2004} studied the velocity effects on the first-order deflection of light and massive particles, based on the Lorentz transformation of harmonic Schwarzschild metric. The method of coordinate transformation \cite{PB1993,Klioner2003,WuckSperh2004} was also employed to discuss the \emph{1PM} motion effects due to both the source of light and the lens in microlensing events~\cite{Heyrovsky2005}. There are many other works devoted to the motion effects appeared in the \emph{1PM} equations of light propagation,
e.g., Refs.~\cite{KopeiMaka2007,ZKS2013,HBL2014b,SH2015}.

These velocity effects, especially relativistic velocity effects, are very likely to be detected nowadays, because almost all their magnitudes are larger than the accuracy of the  high-resolution ($\mu$as level) astronomical programs such as \emph{GAIA mission}~\cite{Perryman2001,Linet2007} and \emph{Japanese Astrometry Satellite Mission for Infrared Exploration (JASMINE)}~\cite{GTKNYM2002,Gouda2004}. Actually, considering the kinematical effects on higher-order gravitational deflection, not limited to first-order deflection, also becomes important. One reason is that the rapid developments of sub-$\mu$as astrometric surveys have been in progress. For example, the planned \emph{Nearby Earth Astrometric Telescope mission (NEAT)}~\cite{MLS2012,MBLJ2014a,Malbet2014b}, which has absorbed some key techniques of the high-accuracy project \emph{Space Interferometry Mission (SIM)}~\cite{Laskin2006,SN2009}, aims at achieving an unprecedented accuracy of $0.05\mu$as. Within the capability of \emph{NEAT}, the nonrelativistic motion effects on the second-order deflection might be detected. When the lens moves quickly with a relativistic velocity, these correctional effects will become so obvious that $\mu$as-level telescope \emph{GAIA} or even \emph{JASMINE} may also detect them. Therefore, it is deserved to investigate the kinematical corrections to the gravitational deflection of test particles up to second post-Minkowskian order (\emph{2PM}). Notice that the discussions of the explicit post-linear equations of motion and light deflection in the field of the two-body system in Ref.~\cite{Brugmann2005} are limited to the low-velocity case.

In the present paper, we investigate the gravitational deflection up to second order of light and (relativistic) neutral massive particles caused by a radially constantly moving Kerr-Newman (KN) balck hole, based on high-accuracy numerical simulations. We restrict our discussions in the weak-field, small-angle, and thin lens approximations. We focus on the kinematically correctional effects induced by the motion of the source on the second-order (leading high-order) deflection. The paper is organized as follows. In Section~\ref{motion-eq}, we start with the harmonic \emph{2PM} metric of the moving KN black hole, and derive the \emph{2PM} equations of motion for test particles via calculating Christoffel symbols. These equations are verified by Euler-Lagrange Method. In Section~\ref{velocityeffects}, the second-order gravitational deflection of light and massive particles are discussed in detail, with the help of numerical calculations. In Section~\ref{application}, we analyze the possibilities of detecting the motion effects on the second-order deflection. Summary is given in Section~\ref{conclusion}. We use units where $G=c=1$ throughout the paper.

\section{Second post-Minkowskian equations of motion for test particles} \label{motion-eq}
Let $\{\bm{e}_1,~\bm{e}_2,~\bm{e}_3\}$ denote the orthonormal basis of a three-dimensional Cartesian coordinate system. We consider a Kerr-Newman black hole with rest mass $M$, electric charge $Q$ and angular momentum $\bm{J}\,(=J\bm{e}_3)$, moving along the positive $x-$axis with a constant velocity vector $\bm{v}~(=v_1\bm{e}_1\equiv v\bm{e}_1)$ (here we only investigate the effects of radial motion of the gravitational source). We denote the rest frame of the background spacetime and the comoving frame of the gravitational source to be $(t,~x,~y,~z)$ and $(X_0,~X_1,~X_2,~X_3)$, respectively. The \emph{2PM} harmonic metric for this moving Kerr-Newman black hole can be written in the coordinate frame $(t,~x,~y,~z)$ as follows~\cite{LinHe2015,LinJiang2014}
{\small\begin{eqnarray}
\hspace*{-70pt}g_{00}=-1+\frac{2\,(1+v^2)\gamma^2M}{R}-\frac{(1+\gamma^2)M^2}{R^2}-\frac{\gamma^2Q^2}{R^2}-\frac{4\,v\gamma^2 a M X_2}{R^3}+\frac{v^2\gamma^2(M^2-Q^2)X_1^2}{R^4}~,
~\label{g00mKN2} \\
\nn \hspace*{-70pt}g_{0i}=\,\gamma\,\zeta_i+v\,\gamma^2\left(-\,\frac{4\,M}{R}+\frac{M^2+Q^2}{R^2}\,\right)\delta_{i1}
-\frac{v\,\gamma\,(\,M^2-Q^2\,)\,X_1\,\left[\,X_i+(\,\gamma-1\,)\,X_1\,\delta_{i1}\,\right]}{R^4} \\
\hspace*{-42pt} +\,\frac{2\left(\gamma^2+v^2\gamma^2-\gamma\right)\,a\,M\,X_2\, \delta_{i1}}{R^3}~,~~~~\label{g0imKN2} \\
\nn \hspace*{-70pt}g_{ij}=\!\left(\!1\!+\!\frac{M}{R}\right)^2\delta_{ij}\!+\!v^2\gamma^2\left(\frac{4M}{R}\!-\!\frac{M^2\!+\!Q^2}{R^2}\right)\delta_{i1}\delta_{j1}
\!-\!v\gamma\left[\zeta_i\,\delta_{j1}\!+\!\zeta_j\,\delta_{i1}\!+\!\frac{4(\gamma\!-\!1)\,aM X_2}{R^3}\,\delta_{i1}\delta_{j1}\right]  \\
\hspace*{-42pt}+\,\frac{\left(\,M^2-Q^2\,\right)\left[\,X_i+(\gamma-1)\,X_1\,\delta_{i1}\,\right]\left[\,X_j+(\gamma-1)\,X_1\,\delta_{j1}\,\right]}{R^4}~, \label{gijmKN2}
\end{eqnarray}}
where $i,~j = 1,~2,~$or $3$, $\gamma= (1-v^2)^{-\scriptstyle \frac{1}{2}}$ is Lorentz factor, and $\delta_{ij}$ denotes Kronecker delta. $\Phi=-\frac{M}{R}$ represents Newtonian gravitational potential, with $\frac{X_1^2+X_2^2}{R^2+a^2}+\frac{X_3^2}{R^2}=1$ and $\mathbf{X}\!\cdot\! d\mathbf{X}\equiv X_1dX_1\!+\!X_2dX_2\!+\!X_3dX_3$.
The symbol $\bm{\zeta}\equiv \frac{2aM}{R^3}\left(\bm{X} \times \bm{e_3}\right)=(\zeta_1,~\zeta_2,~0)$ is a new vector potential~\cite{Weinberg1972} and $a \equiv \frac{J}{M}$ is the angular momentum per mass. We assume the relation $a^2+Q^2\leq M^2$ to avoid the naked singularity for the black hole. In order to calculate the gravitational deflection of test particles up to second order, we only need the following components of the inverse of the metric up to \emph{1PM} order
\begin{eqnarray}
g^{tt}=-1-\frac{2(1+v^2)\gamma^2M}{R}~, \label{g^tt}  \\
g^{xx}=1-\frac{2(1+v^2)\gamma^2M}{R}~, \label{g^xx}   \\
g^{yy}=g^{zz}=1-\frac{2M}{R}~, \label{g^yyzz}  \\
g^{tx}=g^{xt}=-\frac{4v\gamma^2M}{R}~. \label{g^xt}
\end{eqnarray}

Note that the coordinates $X_0,~X_1,~X_2$, and $X_3$ in Eqs.~(\ref{g00mKN2}) - (\ref{gijmKN2}) are related to the coordinates $t,~x,~y$, and $z$ by the common Lorentz transformation
\begin{eqnarray}
X_0=T=\gamma(t-v x)~,  \label{LorentzTran-t} \\
X_1=X=\gamma(x-v t)~,  \label{LorentzTran-x} \\
X_2=Y=y~,  \label{LorentzTran-y} \\
X_3=Z=z~. \label{LorentzTran-z}
\end{eqnarray}
Thus, the partial derivatives of $R$ with respective to $t,~x,~y$, and $z$ can be expressed as
\begin{eqnarray}
\frac{\partial R}{\partial t}=\frac{-v(x-vt)\gamma^2R}{2R^2+a^2-\left[\gamma^2(x-vt)^2+y^2+z^2\right]}~, \label{partial-t}  \\
\frac{\partial R}{\partial x}=\frac{(x-vt)\gamma^2R}{2R^2+a^2-\left[\gamma^2(x-vt)^2+y^2+z^2\right]}~, \label{partial-x}
\end{eqnarray}
\begin{eqnarray}
\frac{\partial R}{\partial y}=\frac{yR}{2R^2+a^2-\left[\gamma^2(x-vt)^2+y^2+z^2\right]}~, \label{partial-y} \\
\frac{\partial R}{\partial z}=\frac{z(R^2+a^2)}{R\left\{2R^2+a^2-\left[\gamma^2(x-vt)^2+y^2+z^2\right]\right\}}~. \label{partial-z}
\end{eqnarray}

For simplicity, we only consider the propagation of test particles confined to the equatorial plane $(z=0)$ of the gravitational lens, and therefore there is one Killing vector field $\left(\frac{\partial }{\partial z}=0\right)$ in the moving KN geometry. After tediously but straightforward calculating the nonvanishing components of the affine connection, as shown in~\ref{A}, we obtain the explicit geodesic equations up to $2PM$ order as follows
{\small\begin{eqnarray}
\hspace*{-70pt}\nn0=\ddot{t}+\frac{v\,\gamma^3\,\dot{t}^{\hspace*{1pt} 2}\,X}{R^2}\left[-\frac{(1+v^2)\,M}{R}+\frac{(v^2-4)\,M^2+Q^2}{R^2}
+\frac{v^2(M^2-Q^2)\,(y^2-X^2)}{R^4}+\frac{6\,v\,aMy}{R^3}\right]  \\
\hspace*{-70pt}\nn+\frac{\gamma^3\,\dot{x}^2X}{R^2}\left\{\frac{v\,(\,v^2\!-\!3\,)\,M}{R}+\frac{v\,[\,(1\!-\!4v^2)\,M^2\!+\!(\,2\!-\!v^2\,)\,Q^2\,]}{R^2}
\!+\!\frac{v\,(M^2\!-\!Q^2)\,(y^2\!-\!X^2)}{R^4}\!+\!\frac{6\,a\,M\,y}{R^3}\right\}  \\
\hspace*{-70pt}\nn+\frac{2\gamma^3\,\dot{t}\,\dot{x}X}{R^2}\!\left[\frac{(1\!+\!v^2)M}{R}\!+\!\frac{3v^2M^2\!-\!Q^2}{R^2}
\!-\!\frac{v^2(M^2\!-\!Q^2)(y^2\!-\!X^2)}{R^4}\!-\!\frac{6\,v a M y}{R^3}\right]\!+\!\frac{2(1\!+\!v^2)\gamma^2M\dot{t}\,\dot{y}\,y}{R^3}  \\
\hspace*{-70pt}-\frac{4\,v\gamma^2\,M\,\dot{x}\,\dot{y}\,y}{R^3}~,\label{geodesic-t}
\end{eqnarray}
\begin{eqnarray}
\hspace*{-70pt}\nn0=\ddot{x}\!+\!\frac{\gamma^3\hspace*{1.5pt}\dot{t}^{\hspace*{1pt} 2}X}{R^2}\!\left[\frac{(1\!-\!3v^2)M}{R}\!+\!\frac{(v^2-4)M^2\!+\!(2v^2\!-\!1)Q^2}{R^2}
+\frac{v^2(M^2\!-\!Q^2)\,(y^2\!-\!X^2)}{R^4}\!+\!\frac{6v^3aMy}{R^3} \right] \\
\hspace*{-70pt}\nn +\frac{\gamma^3\hspace*{1.5pt}\dot{x}^2X}{R^2}\!\!\left[-\frac{(1\!+\!v^2)M}{R}\!+\!\frac{(1\!-\!4v^2)M^2\!+\!v^2Q^2}{R^2}
\!+\!\frac{(M^2\!-\!Q^2)\,(y^2\!-\!X^2)}{R^4}\!+\!\frac{6vaMy}{R^3}\right]\!+\!\frac{2\,v\,\gamma^3\,\dot{t}\,\dot{x}X}{R^2}\!\times\! \\
\hspace*{-70pt} \left[\frac{(1\!+\!v^2)\,M}{R}\!+\!\frac{3M^2\!-\!v^2\,Q^2}{R^2}\!-\!\frac{(M^2\!-\!Q^2)\,(\hspace*{1.5pt}y^2\!-\!X^2\hspace*{1.5pt})}{R^4}
\!-\!\frac{6\,v\,aMy}{R^3}\right]+\frac{2\,\gamma\,M\,(v\dot{X}_0\!-\!\dot{X})\,\dot{y}\,y}{R^3} ~,~ \label{geodesic-x}
\end{eqnarray}
\begin{eqnarray}
\hspace*{-70pt}\nn0=\ddot{y}+\dot{t}^{\hspace*{1pt} 2}
\left\{\frac{\gamma^2\,\,y}{R^2}\left[\frac{(1+v^2\,)\,M}{R}\!-\!\frac{(4+v^2\,)\,M^2+Q^2}{R^2}\!+\!\frac{v^2\,(M^2\!-\!Q^2\,)\,\,(y^2\!-\!X^2)}{R^4}\right]
\!-\!\frac{2\,v\,\gamma^2\,aM}{R^3}\right\}  \\
\hspace*{-70pt}\nn+\,\,\dot{x}^2\left\{\,\frac{\gamma^2\,\,y}{R^2}\left[\,\frac{(\,1+v^2\,)\,M}{R}\!-\!\frac{(\,1+4\,v^2\,)\,M^2+v^2\,Q^2}{R^2}
\!+\!\frac{(\,M^2\hspace*{-1.2pt}-\hspace*{-0.8pt}Q^2\,)\,\,(\,y^2\!-\!X^2\,)}{R^4}\,\right]\!-\!\frac{2\,v\,\gamma^2\,aM}{R^3}\,\right\}  \\
\hspace*{-70pt} +\, 2\,\gamma^2\,\dot{t}\,\dot{x}\!\left[-\frac{2vMy}{R^3}\!+\!\frac{v(5M^2\!+Q^2)y}{R^4}
\!-\!\frac{v(M^2\!-\!Q^2)(y^2\!-\!X^2)y}{R^6}\!+\!\frac{(1\!+\!v^2)aM}{R^3}\right]\!-\!\frac{2M\dot{X}\dot{y}X}{R^3}~, \label{geodesic-y}
\end{eqnarray}}
where dots denote derivatives with respect to $p$ while $p$ is a parameter describing the trajectory, and $\dot{y}$ has been assumed to be the order of $\Phi$. Note that Eqs.~(\ref{geodesic-t}) - (\ref{geodesic-y}) correspond respectively to the $t,~x,$ and $y\,-$component of geodesic equations and that the motion is restricted to the equatorial plane. Eqs.~(\ref{geodesic-t}) - (\ref{geodesic-y}) can also be obtained via Euler-Lagrange Method, as shown in~\ref{B}.
For the case with $v=0$, Eqs.~(\ref{geodesic-t}) - (\ref{geodesic-y}) reduce to the geodesic equations of test particles in the field of a non-moving KN black hole
{\small
\begin{eqnarray}
\hspace*{-70pt}0=\ddot{t}+\frac{2\,\dot{t}\,\left[\,(MR-Q^2)\,\dot{x}\,x+M\,R\,\dot{y}\,y\,\right]}{R^4}+\frac{6\,a\,M\,\dot{x}^2\,x\,y}{R^5}~, \label{geodesic-t-SKN} \\
\hspace*{-70pt}0=\ddot{x}+\frac{\dot{t}^{\hspace*{1pt} 2}x}{R^2}\left(\frac{M}{R}\!-\!\frac{4M^2\!+\!Q^2}{R^2}\right)
\!+\!\frac{\dot{x}^2x}{R^2}\!\left[-\frac{M}{R}\!+\!\frac{M^2}{R^2}\!+\!\frac{(M^2\!-\!Q^2)\,(y^2\!-\!x^2)}{R^4}\right]
\!-\!\frac{2\,M\dot{x}\,\dot{y}\,y}{R^3}~,~\label{geodesic-x-SKN} \\
\nn\hspace*{-70pt}0=\ddot{y}+\frac{\dot{t}^{\hspace*{1pt} 2}y}{R^2}\left(\frac{M}{R}\!-\!\frac{4M^2\!+\!Q^2}{R^2}\right)
+\frac{\dot{x}^2y}{R^2}\left[\frac{M}{R}\!-\!\frac{M^2}{R^2}\!+\!\frac{(M^2\!-\!Q^2)\,(y^2\!-\!x^2)}{R^4}\right]
\!-\!\frac{2M\dot{x}\dot{y}x}{R^3}\!+\!\frac{2aM\dot{t}\dot{x}}{R^3}~,  \\      \label{geodesic-y-SKN}
\end{eqnarray}}
where $R$ reduces to $\sqrt{x^2+y^2-a^2}$ and can be approximated by $\sqrt{x^2+y^2}$ in the computation of the deflection up to second order.
For the Schwarzschild black hole being the gravitational source $(v=a=Q=0)$, Eqs.~(\ref{geodesic-t}) - (\ref{geodesic-y}) are simplified to
{\small
\begin{eqnarray}
\hspace*{-70pt}0=\ddot{t}+\frac{2M\,\dot{t}\,(x\,\dot{x}+y\,\dot{y})}{R^3}~, \label{geodesic-t-SQKN} \\
\hspace*{-70pt}0=\ddot{x}+\frac{M\,x\!\left[\,(R-4M)\,\dot{t}^{\hspace*{1pt} 2}\!-\!(R\!-\!M)\,\dot{x}^2\,\right]}{R^4}\!-\!\frac{2\,M\,\dot{x}\,\dot{y}\,y}{R^3}
+\frac{M^2\,\dot{x}^2\,(y^2-x^2)\,x}{R^6}~,~~~~~~\label{geodesic-x-SQKN} \\
\hspace*{-70pt}0=\ddot{y}+\frac{M\,y \left[(R-4\,M)\,\dot{t}^{\hspace*{1pt} 2}+(R\!-\!M)\,\dot{x}^2\right]}{R^4}\!-\!\frac{2\,M\,\dot{x}\,\dot{y}\,x}{R^3}
\!+\!\frac{M^2\,\dot{x}^2\,(y^2-x^2)\,y}{R^6}~. \label{geodesic-y-SQKN}
\end{eqnarray}}

\section{Gravitational deflection of test particles due to a moving Kerr-Newman black hole} \label{velocityeffects}
In this section, we numerically study the influences of the motion of Kerr-Newman black hole, especially the relativistic velocity effects, on the propagations of test particles including light. We will concentrate on the kinematical corrections to the second-order deflection, since the \emph{1PM} gravitational deflection has been investigated in detail~\cite{WuckSperh2004}.

\subsection{Notations and basics of numerical simulations}  \label{notation}
\begin{figure*}
\begin{center}
  \includegraphics[width=15cm]{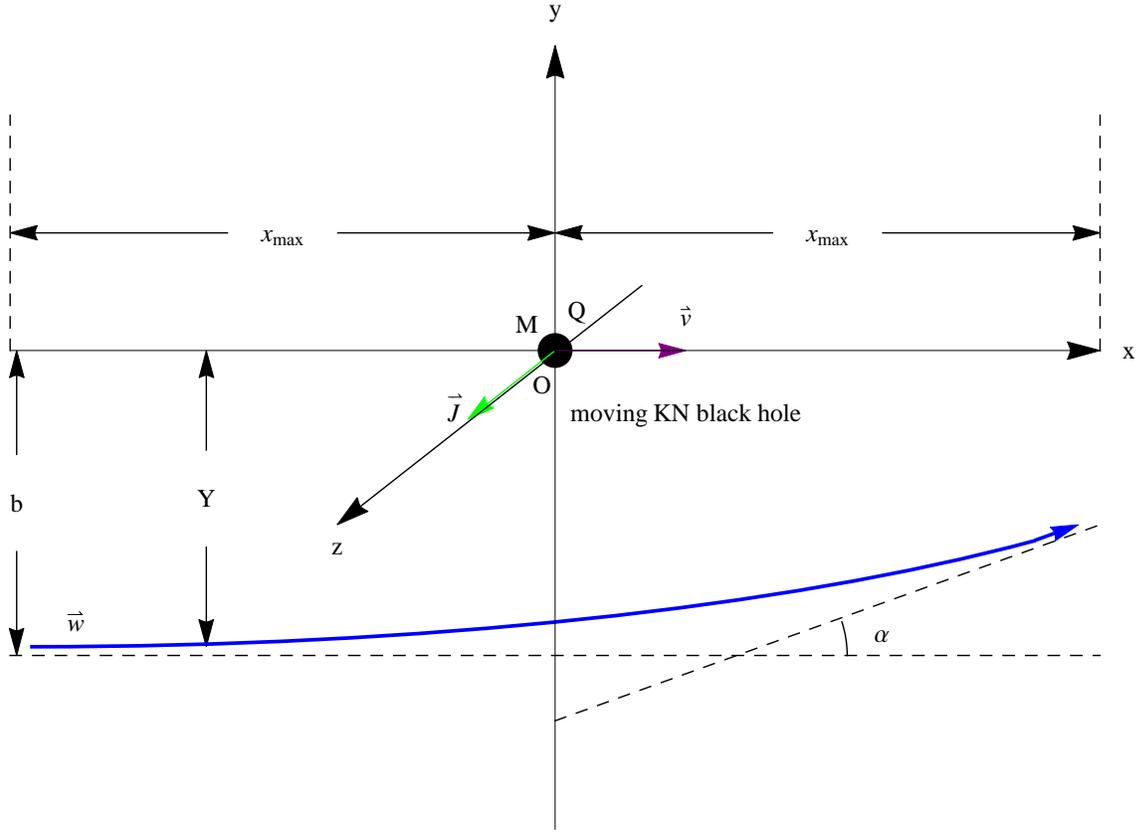}
  \caption{Schematic diagram for gravitational deflection of test particles by a radially moving KN black hole with a constant velocity $\bm{v}=v\bm{e}_1$.
  The starting position of a test particle at the time $t=0$ is assumed to be $(-\infty,-b,~0)$. In the numerical calculations, we use the symbol $-x_{max}$, in which $x_{max}~(>0)$ denotes the finite maximum value that $p$ will take in the simulations, to replace the infinity $-\infty$ (for enough large $x_{max}$).
  The thick blue line represents the propagation path of a test particle coming from $p=-\infty$ with the initial velocity $\bm{w}|_{p\rightarrow-\infty}~(\approx \bm{w}|_{p\rightarrow-x_{max}})~=w\bm{e}_1~$ ($0<w\leq1$ and $v < w$). As mentioned above, the angular momentum vector $\bm{J}$ of the moving source is along positive $z-$axis $(a>0)$, and thus the test particle takes prograde motion relative to the source's spin. The deflection angle $\alpha$ (being greatly exaggerated), detected in the background's rest frame $(t,~x,~y,~z)$, is positive since $Y|_{p\rightarrow-\infty}~(\approx Y|_{p\rightarrow-x_{max}})~=-b<0$. }    \label{Figure1}
\end{center}
\end{figure*}

The initial velocity of a test particle is assumed to be $\bm{w}$, and the impact factor is denoted as $b$. The schematic diagram for gravitational deflection of test particles caused by the moving KN black hole is shown in Fig.~\ref{Figure1}. In order to investigate the kinematically correctional effects, we assume the general form for the gravitational deflection angle of test particles including light up to second order due to the moving KN black hole as
\begin{equation}
\hspace*{-70pt}\alpha(v, \,w)=N_1(v, \,w)\frac{4M}{b}\!+\!N_2(v, \,w)\frac{15\pi}{4}\frac{M^2}{b^2}
\!-\!N_3(v, \,w)\frac{4Ma}{b^2}\!-\!N_4(v, \,w)\frac{3\pi}{4}\frac{Q^2}{b^2} ~, \label{TDAngle0}
\end{equation}
which is based on the analytical formulae in the previous works~\cite{ChakSen2014,ERT2002,ERTprivate,Bhadra2003,EpstShapi1980,EderGod2006}. Here the two-variable function
$N_i(v, \,w)~(i=1,~2,~3,$ or $4)$ represents the kinematical coefficient to characterize the effects of velocities of both the gravitational source and test particle, similar to the definitions in the previous works \cite{WuckSperh2004,LinHe2014}. Notice that for the case of light deflection by a stationary KN black hole ($v=0$ and $w=1$),
$N_1(0, \,1)=N_2(0, \,1)=N_3(0, \,1)=N_4(0, \,1)=1$, and Eq.~(\ref{TDAngle0}) reduces to ~\cite{ChakSen2014}
\begin{equation}
\alpha(0, \,1)=\frac{4M}{b}+\frac{15\pi}{4}\frac{M^2}{b^2}-\frac{4Ma}{b^2}-\frac{3\pi}{4}\frac{Q^2}{b^2} ~. \label{TDAngle1}
\end{equation}

In the numerical simulations, there are six boundary conditions (or starting conditions)
\begin{eqnarray}
\hspace*{-70pt}\left.t(p)\right|_{p=-x_{max}}=-\frac{x_{max}}{w}~,~~~~~~~~\left.x(p)\right|_{p=-x_{max}}=-x_{max}~,~~~~~~~~\left.y(p)\right|_{p=-x_{max}}=-b ~,~~~~ \label{initial-condition-1}  \\
\hspace*{-70pt}\left.\dot{t}(p)\right|_{p=-x_{max}}=\frac{1}{w}~,~\hspace*{42pt}~\left.\dot{x}(p)\right|_{p=-x_{max}}=1~,
~\hspace*{46pt}~\left.\dot{y}(p)\right|_{p=-x_{max}}=0~. \label{initial-condition-2}
\end{eqnarray}
The computation domain for the trajectory parameter is set as $p \in [-x_{max},~x_{max}]$. The value of the parameter $b$ is chosen as $1.0\times10^{5}M$ to guarantee a weak field, and $x_{max}$ is much larger than $b$ and chosen as $1.0\times10^{10}M~(=1.0\times10^{5}b)$. The mass $M$ of the gravitational source is set as
$2.5\times10^{6}M_{\odot}~(\sim 3.6875\times10^6 km)$ which is close to the mass of Sagittarius A$^*$ at the galactic center~\cite{BS1999,HRRTCM1996,EckaGen1997,Narayan1998}, where $M_{\odot}$ is the mass of the sun. Notice that the conclusions below are independent on this specifically chosen value of the gravitational mass. The deflection angle of a test particle can be numerically calculated via integrating the geodesic equations (i.e., Eqs.~(\ref{geodesic-t}) - (\ref{geodesic-y})) as follow
\begin{equation}
\alpha(v,~w)_{N}=\left(\arctan{\dot{y}}\right)_{p\rightarrow x_{max}}=\left.\arctan{\frac{dy}{dp}}\right|_{p\rightarrow x_{max}}~. \label{Numerical-Angle}
\end{equation}
Here and thereafter, the quantity with the subscript $N$ denotes the value obtained by the numerical calculation. We employ Mathematica to do all calculations,
in which $AccuracyGoal=39$ and $PrecisionGoal=13$ are chosen and the numerical methods {\tt \emph{NDSolve} } and {\tt \emph{ParametricNDSolve}} for solving the equations of motion are used.

\subsection{Gravitational deflection of light up to second order}
For the light deflection, the coefficient $N_i(v, \,w)$ in Eq.~(\ref{TDAngle0}) reduces to $N_i(v, \,1)$, and the analytical forms for the first and third coefficients have been given in previous works~\cite{WuckSperh2004,LinHe2014,Sereno2005}: $N_1(v, \,1)=N_3(v, \,1)=(1-v)\gamma$. Therefore, we concentrate on $N_2(v, \,1)$ and $N_4(v, \,1)$, corresponding to the contributions by the second-order moving-Schwarzschild deflection and the charge-induced deflection respectively.

\subsubsection{Determination of kinematical coefficient $N_2(v, \,1)$}  \label{N2}
\begin{figure*}
\begin{center}
  \includegraphics[width=12cm]{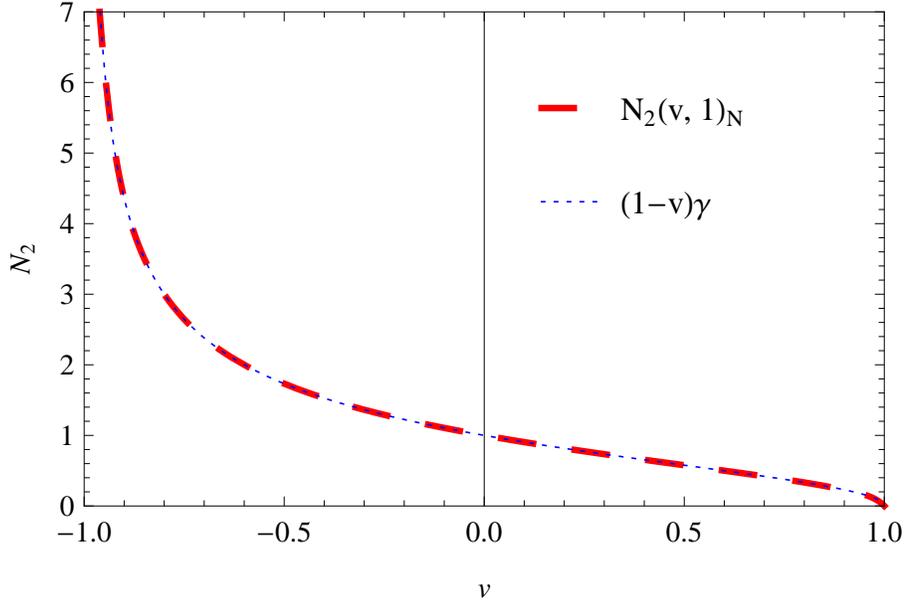}
  \caption{Comparison between the numerical result (long dashed red line) of $N_2(v, \,1)$ and the analytical coefficient $N_1(v, \,1)=(1-v)\gamma$ (short dashed blue line). }    \label{Figure2}
\end{center}
\end{figure*}
The second-order Schwarzschild deflection $\frac{15\pi}{4}\frac{M^2}{b^2}$ is larger than the second-order Kerr term $\frac{4Ma}{b^2}$, hence the kinematical correctional effect on the former is very likely to be more obvious than that on the latter. In order to determine $N_2(v, \,1)$, we can numerically integrate
Eqs.~(\ref{geodesic-t}) - (\ref{geodesic-y}) with $a=Q=0$, utilize the explicit form of $N_1(v, \,1)$, and finally express $N_2(v, \,1)_N$ as follow
\begin{equation}
N_2(v, \,1)_N=\frac{\alpha(v, \,1)_{N-SS}-\frac{4(1-v)\gamma M}{b}}{\frac{15\pi M^2}{4b^2}}~, \label{N2-1}
\end{equation}
where $\alpha(v, \,1)_{N-SS}$ denotes the numerical result of light deflection angle up to second order due to a moving-Schwarzschild source.

Fig.~\ref{Figure2} presents the numerical result of $N_2(v, \,1)$ with various velocity $v$. For comparison $N_1(v, \,1)$ is also plotted in the figure. We surprisedly find that
$N_2(v, \,1)_N$ is consistent with $N_1(v, \,1)$ and the relative error is less than $0.01\%$. In other words, the kinematical coefficient in the second-order moving-Schwarzschild contribution is the same as that in the first-order term, i.e., $N_2(v, \,1)=N_1(v,~1)=(1-v)\gamma$.

\subsubsection{Determination of kinematical coefficient $N_4(v, \,1)$}    \label{N4}
\begin{figure*}[t]
  \setlength{\unitlength}{1cm}
\begin{center}
  \begin{minipage}[b]{12.1cm}
  \centering
  \includegraphics[width=12.1cm]{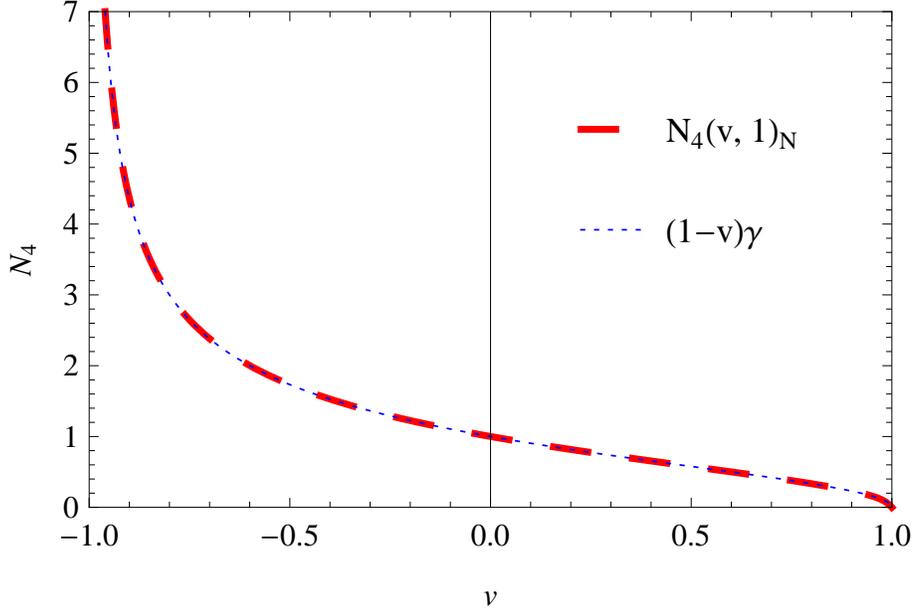}
  \end{minipage}
   \caption{$N_4(v, \,1)_N$ (long dashed red line) plotted to compare with $N_1(v, \,1)=N_2(v, \,1)=(1-v)\gamma$ (short dashed blue line), with $Q=0.99M$. Notice that here $Q$ should be chosen as large as possible to reduce the computational error, though any value in the range $(0,M)$ can be chosen theoretically.  } \label{Figure3}
\end{center}
\end{figure*}
The charge $Q$ of the black hole can also induce a gravitational deflection of test particle~\cite{ERT2002}. Similarly, we determine $N_4(v, \,1)$ by numerically solving the \emph{2PM} geodesic equations of light in the field of a moving Reissner-Nordstr\"{o}m (RN) black hole, i.e., integrating Eqs.~(\ref{geodesic-t}) - (\ref{geodesic-y}) with $a=0$. Based on the explicit forms of $N_1(v, \,1)$ and $N_2(v, \,1)$, the numerical result of $N_4(v, \,1)$ can be expressed as
\begin{equation}
N_4(v,~\!1)_N=\frac{\alpha(v,~1)_{N-RN}-(1-v)\gamma\left(\frac{4M}{b}+\frac{15\pi}{4}\frac{M^2}{b^2}\right)}{-\frac{3\pi Q^2}{4b^2}}~, \label{N4-1}
\end{equation}
where $\alpha(v,~\!1)_{N-RN}$ denotes the numerical result of light deflection angle up to the second order caused by the moving RN source.

Considering the fact that $N_1(v, \,1)=N_2(v, \,1)=(1-v)\gamma$, we conjecture $N_4(v, \,1)$ might also be $(1-v)\gamma$. Fig.~\ref{Figure3} shows the comparison between $N_4(v, \,1)_N$ and $(1-v)\gamma$, and it can be seen that they match with each other perfectly. Therefore we have $N_4(v, \,1)=(1-v)\gamma$.

\subsubsection{Light deflection angle up to second order}
\begin{table}[t]
\scriptsize
\begin{center}
\begin{tabular}{ccccc}
        \hline
       $v$ & $\alpha(v, \,1)_N~(rad)$ & $\alpha(v, \,1)~(rad)$ & $\xi_{total} (\%)$       \\
       \hline
       0.9 &   0.000009176840208    &  0.000009176840232   & 2.62$\times10^{-7}$      \\
       0.5 &   0.000023094541442    &  0.000023094541464   & 9.22$\times10^{-8}$      \\
       0.1 &   0.000036182192760    &  0.000036182192790   & 8.39$\times10^{-8}$      \\
      0.01 &   0.000039602890160    &  0.000039602890194   & 8.77$\times10^{-8}$      \\
     0.001 &   0.000039960938218    &  0.000039960938254   & 8.84$\times10^{-8}$      \\
 $0.00001$ &   0.000040000519150    &  0.000040000519185   & 8.84$\times10^{-8}$      \\
         0 &   0.000040000919157    &  0.000040000919192   & 8.82$\times10^{-8}$      \\
$-0.00001$ &   0.000040001319168    &  0.000040001319204   & 8.84$\times10^{-8}$      \\
  $-$0.001 &   0.000040040940096    &  0.000040040940132   & 8.88$\times10^{-8}$      \\
   $-$0.01 &   0.000040402948546    &  0.000040402948582   & 8.94$\times10^{-8}$      \\
    $-$0.1 &   0.000044222680035    &  0.000044222680077   & 9.57$\times10^{-8}$      \\
    $-$0.5 &   0.000069283624270    &  0.000069283624391   & 1.75$\times10^{-7}$      \\
    $-$0.9 &   0.000174359962643    &  0.000174359964408   & 1.01$\times10^{-6}$      \\
        \hline
\end{tabular}\par
  \caption{The relative difference between the analytical and numerical results for the light deflection angle. $\alpha(v, \,1)_N$ is defined in Eq.~(\ref{Numerical-Angle}). $\xi_{total}$ denotes the relative difference (or relative error) between them. As an example, here we set $a=Q=0.5M$ in the numerical simulation.}  \label{Table1}
\end{center}
\end{table}

\begin{figure*}
\begin{center}
  \includegraphics[width=12.72cm]{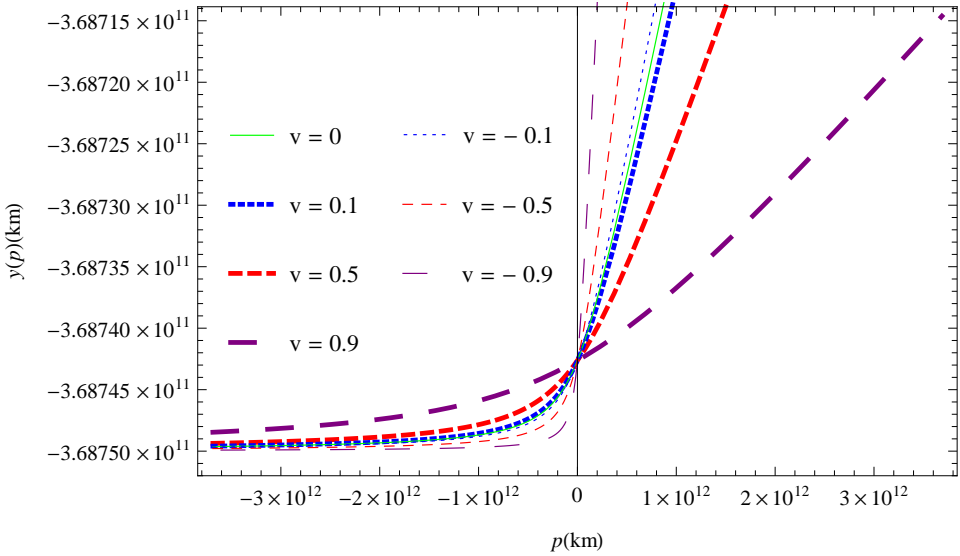}
  \caption{The trajectories of light in the time-dependent gravitational field of the moving KN black hole for various $v$, with $a=Q=0.5M$ (as an example).} \label{Figure4}
\end{center}
\end{figure*}

From the discussions above, the light deflection angle up to second order due to a constantly radially moving Kerr-Newman black hole can be written as
\begin{equation}
\alpha(v, \,1)=(1-v)\gamma\left(\frac{4M}{b}+\frac{15\pi}{4}\frac{M^2}{b^2}-\frac{4Ma}{b^2}-\frac{3\pi}{4}\frac{Q^2}{b^2}\right)~. \label{TDAngle2}
\end{equation}
It should be emphasized that this formula is obtained based on numerical calculations and still needs to be confirmed by an analytical calculation. Table~\ref{Table1} gives the comparison between the analytical and numerical results for the moving KN deflection angle of light, and we can see their difference is very small (average difference is about $0.01\mu$as). Fig.~\ref{Figure4} shows the propagation paths of light in the time-dependent field of the moving KN black hole for various $v$. It can be seen that the correctional effects become distinguishable when the velocity $v$ of the moving source is relativistic such as $|v|\gtrsim 0.02$, and turn to be very obvious for the highly relativistic motion (e.g., $|v|>0.5$).

Eq.~(\ref{TDAngle2}) indicates that the kinematically correctional factor $(1-v)\gamma$ applies not only to the first-order gravito-electric deflection~\cite{WuckSperh2004},
but also to the second-order gravitational deflection, including the second-order gravito-electric, gravito-magnetic, and charge-induced deflections. In the limit of
low velocity ($|v|\ll1$), Eq.~(\ref{TDAngle2}) reduces to
\begin{equation}
\hspace*{-70pt}\alpha(v, \,1)=\frac{4M}{b}\!+\!\frac{15\pi}{4}\frac{M^2}{b^2}\!-\!\frac{4Ma}{b^2}\!-\!\frac{3\pi}{4}\frac{Q^2}{b^2}
\!-v\left(\frac{4M}{b}\!+\!\frac{15\pi}{4}\frac{M^2}{b^2}\!-\!\frac{4Ma}{b^2}\!-\!\frac{3\pi}{4}\frac{Q^2}{b^2}\right)~,~\label{TDAngle2-2}
\end{equation}
which extends the previous kinematically correctional result obtained in the \emph{FOV} and \emph{FOD} approximations~\cite{PB1993,WuckSperh2004,FKN2002,MB2003}
\begin{equation}
\alpha(v, \,1)=\frac{4M}{b}-\frac{4vM}{b}~. \label{TDAngle2-3}
\end{equation}

\subsection{Gravitational deflection of massive particle up to second order}
In this section, we investigate the gravitational deflection of massive particles up to second order, and discuss the kinematical effects on the massive particle deflection due to the moving KN black hole. The small-angle and weak-field approximations are used to restrict the initial velocity $w\in[w_{min},~1)$, where $w_{min} (>0)$ denotes the lower limit of $w$ and depends on the impact factor $b$. For example, $w$ should satisfy the condition $w\gtrsim 0.36$, supposing ${b}$ and the small deflection angle $\alpha$ are set to be $1.0\times10^{5}M$ and $0.01^{\circ}$, respectively.

\subsubsection{Massive particle deflection by a non-moving Schwarzschild black hole}

In the literature, there exist two different analytical formulae for Schwarzschild deflection of massive particle up to second order as follow~\cite{AccRagu2002,BSN2007}
\begin{eqnarray}
\hspace*{-70pt}\alpha(0, \,w)_{AR}=2\left(1+\frac{1}{w^2}\right)\frac{M}{b}+3\pi\left(\frac{1}{4}+\frac{1}{w^2}\right)\frac{M^2}{b^2}~,\\
\hspace*{-70pt}\alpha(0, \,w)_{BSN}=2\left(1+\frac{1}{w^2}\right)\frac{M}{b}+\left[3\pi\left(\frac{1}{4}+\frac{1}{w^2}\right)+2\left(1-\frac{1}{w^4}\right)\right]\frac{M^2}{b^2}~,
\end{eqnarray}
where $\alpha(0, \,w)_{AR}$ and $\alpha(0, \,w)_{BSN}$ denote the analytical formulations given in Refs.~\cite{AccRagu2002} and~\cite{BSN2007}, respectively. The difference is in the second-order terms. Here we can numerically solve the \emph{2PM} geodesic equations of massive particle in the Schwarzschild spacetime, i.e., Eqs.~(\ref{geodesic-t-SQKN}) - (\ref{geodesic-y-SQKN}), to examine the reported formulae.

\begin{table}[t]
\scriptsize
\begin{center}
   \begin{tabular}{cccccc}
        \hline
           w & $\alpha(0, \,w)_{N}$ & $\alpha(0, \,w)_{AR}$ & $\alpha(0, \,w)_{BSN}$ &   $\xi_{AR}$ (\%)   &   $\xi_{BSN}$ (\%)     \\
       \hline
        0.99 &  0.000040407278237    &  0.000040407278246   &   0.000040407270041    & 2.12$\times10^{-8}$ & 2.03$\times10^{-5}$    \\
         0.9 &  0.000044692757184    &  0.000044692757197   &   0.000044692652365    & 2.96$\times10^{-8}$ & 2.35$\times10^{-4}$    \\
         0.5 &  0.000100004005252    &  0.000100004005531   &   0.000100001005531    & 2.79$\times10^{-7}$ & 3.00$\times10^{-3}$    \\
        0.36 &  0.000174328493448    &  0.000174328495479   &   0.000174316787995    & 1.16$\times10^{-6}$ & 6.72$\times10^{-3}$    \\
        \hline
   \end{tabular}\par
    \caption{The comparison among numerical and theoretical Schwarzschild deflection angles of a massive particle up to second order. $\xi_{AR}$ and $\xi_{BSN}$ denote the relative errors of $\alpha(0, \,w)_{AR}$ and $\alpha(0, \,w)_{BSN}$ with respect to the numerical result $\alpha(0, \,w)_{N}$, respectively. }   \label{Table2}
\end{center}
\end{table}

\begin{figure*}
\begin{center}
  \includegraphics[width=12.65cm]{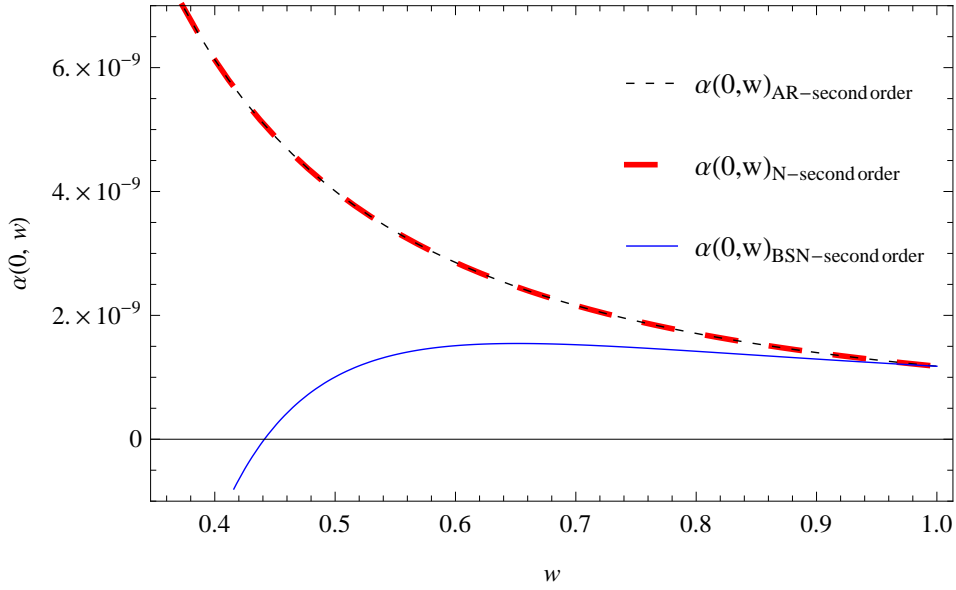}
  \caption{ The comparison among the second-order contributions given by $\alpha(0, \,w)_{AR}$, $\alpha(0, \,w)_{BSN}$, and $\alpha(0, \,w)_{N}$, respectively.}    \label{Figure5}
\end{center}
\end{figure*}

Table~\ref{Table2} presents the comparison among $\alpha(0, \,w)_{AR}$, $\alpha(0, \,w)_{BSN}$, and the numerical result $\alpha(0, \,w)_N$ for various velocity $w$ of massive particle. We can see that $\alpha(0, \,w)_{AR}$ agrees with $\alpha(0, \,w)_{N}$ much better than $\alpha(0, \,w)_{BSN}$ does. The difference is small since the first-order terms are dominant. Fig.~\ref{Figure5} shows the comparison among the second-order contributions given by $\alpha(0, \,w)_{AR}$, $\alpha(0, \,w)_{BSN}$, and $\alpha(0, \,w)_{N}$, respectively. We can see that $\alpha(0, \,w)_N$ matches with $\alpha(0, \,w)_{AR}$ perfectly, and differs from $\alpha(0, \,w)_{BSN}$. Note that the numerical result is based on the harmonic Schwarzschild metric, which is different from the approaches in the previous works.

\subsubsection{Massive particle deflection by a non-moving Kerr black hole}

Based on the analytical formulations of Schwarzschild deflection of massive particle~\cite{AccRagu2002} and the second-order Kerr contribution~\cite{LinHe2014}, we can write down the deflection angle of a massive particle up to second order due to a stationary Kerr black hole as
\begin{equation}
\alpha(0, \,w)=2\left(1+\frac{1}{w^2}\right)\frac{M}{b}+3\pi\left(\frac{1}{4}+\frac{1}{w^2}\right)\frac{M^2}{b^2}-\frac{1}{w}\frac{4Ma}{b^2}~.\label{KERR-mass}
\end{equation}

\begin{figure*}
\begin{center}
  \includegraphics[width=11.93cm]{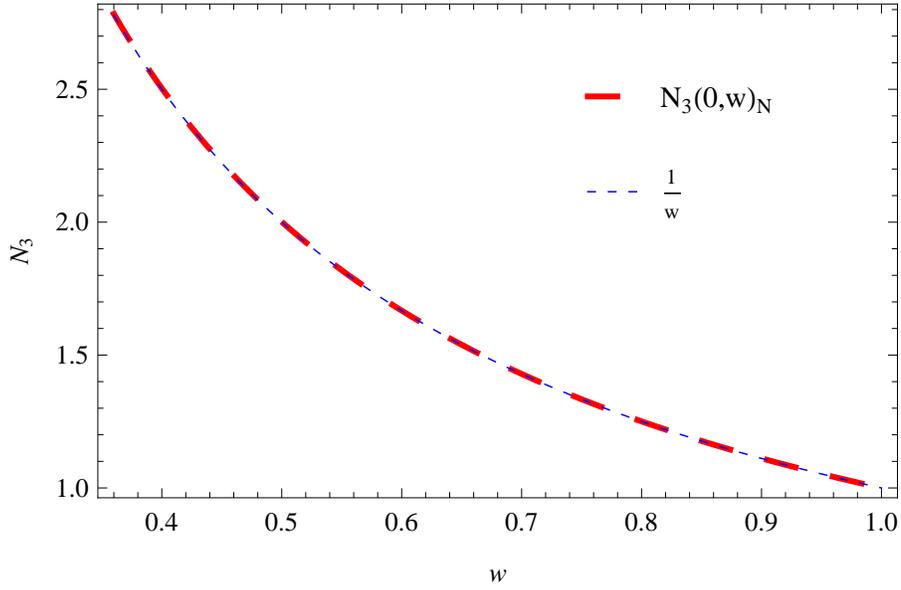}
  \caption{The comparison between the numerical result (long dashed red line) of the coefficient $N_3(0, \,w)$ and its theoretical value (short dashed blue line), with $a=0.99M$ as an example. }    \label{Figure6}
\end{center}
\end{figure*}

Fig.~\ref{Figure6} presents the comparison between the analytical coefficient $N_3(0, \,w)=1/w$ and its numerical computation which is defined as
\begin{equation}
\hspace*{-70pt}N_3(0, \,w)_N=\frac{\left.\arctan{\frac{\partial y(w,~p)}{\partial p}}\right|_{p\rightarrow x_{max}} -2\left(1+\frac{1}{w^2}\right)\frac{M}{b}-3\pi\left(\frac{1}{4}+\frac{1}{w^2}\right)\frac{M^2}{b^2}}{-\frac{4Ma}{b^2}}~.   \label{N-staticKerr-N3}
\end{equation}
It can be seen that the theoretical value of $N_3(0, \,w)$ matches with the numerical result very well.

\subsubsection{Massive particle deflection by a non-moving KN black hole}

\begin{figure*}
\begin{center}
  \includegraphics[width=11.6cm]{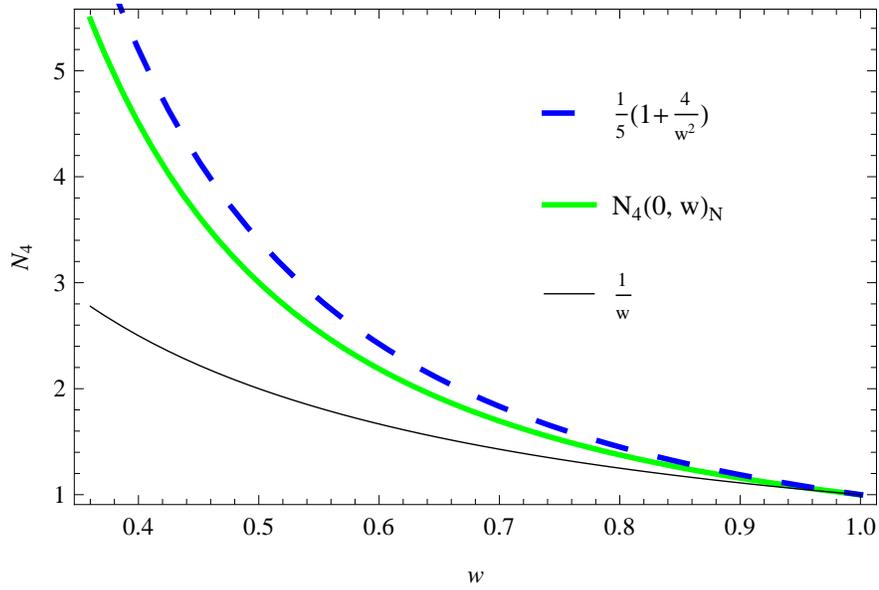}
  \caption{$N_4(0, \,w)_N$ (thick green line) plotted to compare with the analytical coefficients $N_2(0, \,w)$ (dashed blue line) and $N_3(0, \,w)$ (thin black line). As an example, here we set $a=0.1M$ and $Q=0.99M$. }  \label{Figure7}
\end{center}
\end{figure*}

The second-order charge-induced contribution to gravitational deflection of massive particle is characterized by the term $N_4(0, \,w)\frac{3\pi}{4}\frac{Q^2}{b^2}$. We can solve Eqs.~(\ref{geodesic-t-SKN}) - (\ref{geodesic-y-SKN}) to obtain the numerical value of the coefficient $N_4(0, \,w)$. Fig.~\ref{Figure7} shows $N_4(0, \,w)_N$ as the function of $w$. For comparison, $N_2(0, \,w)=\frac{1}{5}\left(1+\frac{4}{w^2}\right)$ and $N_3(0, \,w)=\frac{1}{w}$ are also given.

\subsubsection{Massive particle deflection by a moving KN black hole}
\begin{figure*}
\begin{center}
  \includegraphics[width=14cm]{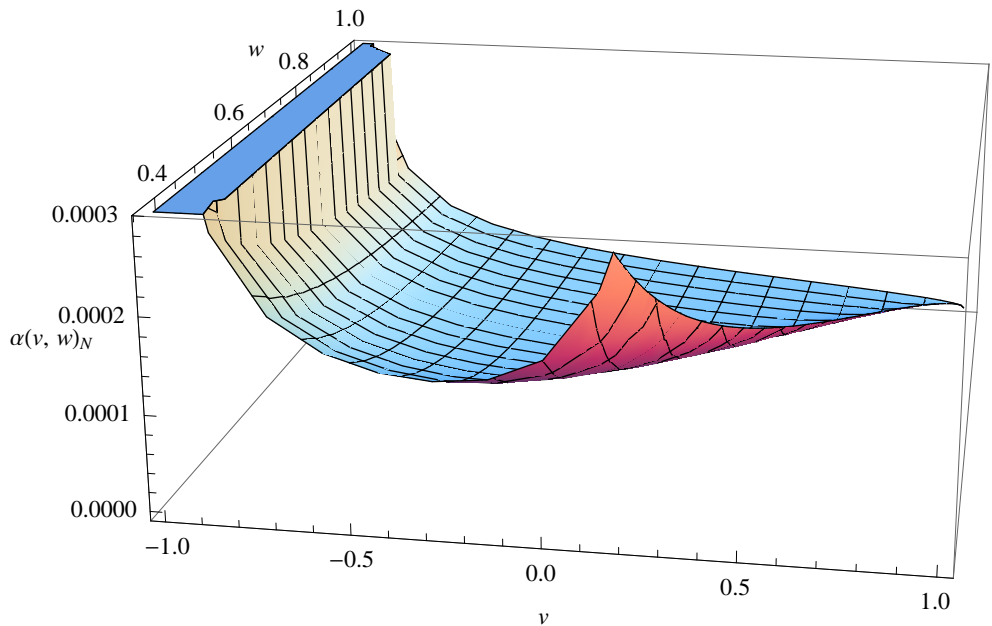}
  \caption{$\alpha(v, \,w)_N$ plotted as the function of two variables $v$ and $w$, with $a=Q=0.5M$.
  Here we set the range of $v$ to be $-0.999\leq v\leq0.999~(v<w)$ in the simulations. }    \label{Figure8}
\end{center}
\end{figure*}

For a general case $(a\neq0,~Q\neq0,~-1<v<1,~w_{min}\leq w<1)$, we can also numerically solve Eqs.~(\ref{geodesic-t}) - (\ref{geodesic-y}) and utilize Eq.~(\ref{Numerical-Angle}) to calculate the gravitational deflection angle $\alpha(v, \,w)_N$ up to second order. Fig.~\ref{Figure8} presents $\alpha(v, \,w)_N$ as the function of both $v$ and $w$ for the moving Kerr-Newman black hole with $a=Q=0.5M$ as an example.

\vspace{20pt}
Up to now, we have discussed the gravitational deflection of test particles up to second order by the moving KN source, with the help of numerical simulations. In the next section, we will analyze the possibilities to detect the kinematical corrections to the second-order deflection in the astronomical observations.

\section{Possibilities to detect the kinematically correctional effects} \label{application}
The techniques of high-accuracy angle measurement at the level of $\mu$as have been achieved in nowadays astronomical projects. The accuracy of ESA's telescope \emph{GAIA}~\cite{Gaia2015} is about $7\mu$as and $25\mu$as for V magnitude $=12$ and $15$, respectively~\cite{MCM2007}. In contrast to \emph{GAIA}, the proposed project \emph{SIM} had achieved a higher accuracy $\sim1\mu$as for narrow angle though it was cancelled. As mentioned above, \emph{NEAT} plans to achieve a much higher accuracy than \emph{SIM}. These high-accuracy astronomical surveys, greatly promote the theoretical investigations of the detectable kinematical effects (especially relativistic motion effects) which appear in leading high-order terms of classic tests of general relativity, such as time delay~\cite{HBL2014a} and frequency shift~\cite{HBL2014a,HBL2012} of electromagnetic waves. The theoretical model established in this work, shows a possibility to observe the velocity effects on the second-order gravitational deflection.

We first consider light deflection. Table~\ref{Table3} gives the magnitude of the kinematical correction
$\Delta(v, \,a, \,Q)=\left[1-(1\!-\!v)\gamma\right]\left(\frac{15\pi}{4}\frac{M^2}{b^2}\!-\!\frac{4Ma}{b^2}\!-\!\frac{3\pi}{4}\frac{Q^2}{b^2}\right)$ to the second-order deflection given in Eq.~(\ref{TDAngle2}). It is found that the correctional effect $\Delta(v, \,a, \,Q)$ on the second-order deflection may be larger than the accuracy of \emph{NEAT}, even though the source is nonrelativistic (not to mention the relativistic case). For example, when the velocity of a moving Schwarzschild source ($a=Q=0$) is about $2.058\times10^{-4}\sim61.7 km/s$, the kinematical correction to the second-order deflection angle will reach $0.05\mu$as. This velocity is lower than the velocities of many celestial bodies, such as star $\mu$ Cas (space velocity $\sim145 km/s$)~\cite{KK1953}, X-ray point source RX J$0822-4300$ (recoil velocity $>500 km/s$)~\cite{HB2006,BPWP2012}, and pulsars B$2224+65$ (transverse velocity $\geq800 km/s$)~\cite{CRL1993} and B$1508+55$ (transverse velocity $\sim1083^{+103}_{-90} km/s$)~\cite{CVB2005}. The heliocentric radial velocity ($\sim620 km/s$) of the first hypervelocity star in \emph{Large Sky Area Multi-Object Fiber Spectroscopic Telescope} \emph{(LAMOST)} survey~\cite{ZZCP2012,CZC2012,Zheng2014} is much larger than this velocity. Even the circular velocity ($\sim220 km/s$) of the sun around galaxy center also exceeds it. Therefore, there is a good possibility to detect the nonrelativistic kinematical effects on the second-order light deflection by high-accuracy telescopes such as \emph{NEAT}.
\begin{table}[t]
\scriptsize
\begin{center}
   \begin{tabular}{cccccccc}
        \hline
        v~$\setminus$~(\,a,~Q\,) & (\,0.99M,~0\,) & (\,0.99M,~0.1M\,) & (\,0.5M,~0.5M\,) & (\,0.1M,~0.99M\,) & (\,0,~0.99M\,) & (\,0,~~0\,)     \\
       \hline
            0.9                  &   124.31       &   123.94          &    146.10        &   144.19          &   150.55       &   187.25        \\
            0.1                  &   15.40        &   15.35           &    18.10         &   17.86           &   18.65        &   23.20         \\
           0.01                  &   1.61         &   1.60            &    1.89          &   1.86            &   1.94         &   2.42          \\
          0.001                  &   0.16         &   0.16            &    0.19          &   0.19            &   0.20         &   0.24          \\
         0.0001                  &   $\star$      &   $\star$         &    $\star$       &   $\star$         &   $\star$      &   $\star$       \\
       \hline
   \end{tabular}\par
\caption{The magnitude ($\mu$as) of the kinematical correction $\Delta(v, \,a, \,Q)$ to the second-order light deflection for various $v$. Several combinations of $a$ and $Q$ are listed as examples. The star ``$\star$" denotes the value which is less than $0.05\mu$as (the accuracy of \emph{NEAT}). Notice that here we present the cases with high-magnitude charge mainly for illustration, since in most cases the possible original charge of the black hole in the Universe may have been neutralized or become very small up to now.  }   \label{Table3}
\end{center}
\end{table}

We then consider a massive particle with a relativistic initial velocity as the test particle, such as a high-speed neutron in secondary cosmic rays. We can numerically estimate the magnitude of the kinematical correction $\Delta(v, \,w, \,a, \,Q)$ to the second-order massive particle deflection. We take a neutron with $w\!=\!0.5$ and set $a\!=\!Q\!=\!0.5M$ as an example. It is found that $\Delta(v, \,0.5, \,a, \,Q)$ is about $94.68 \mu$as,~$0.70 \mu$as,~$0.07 \mu$as ($>0.05 \mu$as) for $v=0.1,~0.001,~0.0001$ ($\sim 30km/s$), respectively. We conclude that the possibility to detect the nonrelativistic correctional effects on the second-order deflection of massive particle via these high-accuracy telescopes is also large.

\section{Summary} \label{conclusion}

In this paper, starting from the \emph{2PM} harmonic metric of the radially moving KN black hole, we have derived the explicit equations of motion and investigated the gravitational deflection of test particles including light up to second order, based on the high-accuracy numerical calculations. We focus on discussing the detectable kinematical effects
(including both relativistic and nonrelativistic correctional effects) on the second-order deflection.

Main results are summarized as follows. Firstly, we obtain the analytical form for the gravitational deflection angle of light up to second order due to the moving KN source (see Eq.~(\ref{TDAngle2})). Secondly, our numerical calculations verify the analytical formula (given in Ref.~\cite{AccRagu2002}) for the Schwarzschild deflection of a massive particle up to second order. Thirdly, the analytical massive particle deflection angle up to second order in the Kerr geometry is achieved (see  Eq.~(\ref{KERR-mass})). Fourthly, our numerical approach can be used to calculate the deflection angle of a massive particle up to second order due to the moving KN source. Finally, the possibilities for detecting the kinematical effects by the high-resolution astronomical surveys such as \emph{NEAT} are also discussed.

\section*{ACKNOWLEDGEMENT}
We would like to thank the anonymous reviewers for their constructive comments and suggestions on improving the quality of this paper. This work was supported in part by the National Natural Science Foundation of China (Grant No. 11547311), the National Basic Research Program of China (Grant No. 2013CB328904) and the Fundamental Research Funds for the Central Universities (No. 2682014ZT32).

\appendix

\section{nonvanishing components of Christoffel symbols} \label{A}

Based on Eqs.~(\ref{g00mKN2}) - (\ref{partial-z}), we directly derive the nonvanishing Christoffel symbols as follows
{\small
\begin{eqnarray}
\hspace*{-70pt}&&\Gamma^t_{tt}=\frac{v\,\gamma^3X}{R^2}\!\left[-\frac{(\hspace*{0.8pt}1\!+\!v^2\hspace*{0.9pt})\,M}{R}\!-\!\frac{(4-v^2)\,M^2-Q^2}{R^2}
\!+\!\frac{v^2\,(M^2\!-\!Q^2)\,(y^2\!-\!X^2)}{R^4}\!+\!\frac{6\,v\,aMy}{R^3}\right]~,~\label{ttt} \\
\hspace*{-70pt}&&\Gamma^t_{tx}=\Gamma^t_{xt}=\frac{\gamma^3\,X}{R^2}\left[\frac{(\,1\!+\!v^2\hspace*{0.9pt})\,M}{R}+\frac{3\,v^2M^2\!-Q^2}{R^2}
\!-\!\frac{v^2\,(M^2\!-\!Q^2)\,(y^2\!-\!X^2)}{R^4}\!-\!\frac{6\,v\,aMy}{R^3}\right] ~,~~~~~~~~\label{ttx}  \\
\hspace*{-70pt}&&\Gamma^t_{ty}=\Gamma^t_{yt}=\!\frac{\gamma^2\,y}{R^2}\!\left[\frac{(1\!+\!v^2)M}{R}\!-\!\frac{v^2M^2\!+\!Q^2}{R^2}
\!+\!\frac{2\,v^2(M^2\!-\!Q^2)X^2}{R^4}\!-\!\frac{6\hspace*{1pt}v\hspace*{1pt}aMy}{R^3}\right]\!+\!\frac{2\hspace*{0.9pt}v\hspace*{0.9pt}\gamma^2\hspace*{1pt}aM}{R^3} ~,  \label{tty} \\
\hspace*{-70pt}&&\Gamma^t_{xy}\!=\!\Gamma^t_{yx}\!=-\gamma^2\!\left\{\!\frac{vy}{R^2}\!\left[\frac{2M}{R}\!-\!\frac{M^2\!+\!Q^2}{R^2}\!+\!\frac{2(M^2\!-\!Q^2)X^2}{R^4}\right]
\!+\!\frac{aM\left[\hspace*{1pt}3(X^2\!-\!y^2)\!-\!v^2R^2\hspace*{1pt}\right]}{R^5}\!\right\}~,  \label{txy} \\
\hspace*{-70pt}&&\Gamma^t_{xx}\!=\!\frac{\gamma^3\!X}{R^2}\!\!\left[\!\frac{v(v^2\!\!-\!3)M}{R}\!+\!\frac{v[(1-4v^2)M^2\!+\!(2\!-\!v^2)Q^2]}{R^2}
\!+\!\frac{v(M^2\!\!-\!Q^2)(y^2\!\!-\!\!X^2)}{R^4}\!+\!\frac{6aMy}{R^3}\!\right]~,   \label{txx} \\
\hspace*{-70pt}&&\Gamma^t_{yy}=\frac{\gamma\,X}{R^2}\left[\frac{v\,M}{R}-\frac{vM^2}{R^2}\!+\!\frac{v\,(M^2\!-\!Q^2)\,(X^2\!-\!y^2)}{R^4}\!-\!\frac{6\,a\,My}{R^3}\right]~,~~~~\label{tyy} \\
\hspace*{-70pt}&&\Gamma^x_{tt}\!=\!\frac{\gamma^3\!X}{R^2}\hspace*{-2.5pt}\left[\!\frac{(1\!-\!3v^2)M}{R}\!+\!\frac{(v^2\!-\!4)M^2\!+\!(2v^2\!-\!1)Q^2}{R^2}
\!+\!\frac{v^2(M^2\!\!-\!Q^2)(y^2\!\!-\!\!X^2)}{R^4}\!+\!\frac{6v^3aMy}{R^3}\!\right]~,    \label{xtt}  \\
\hspace*{-70pt}&&\Gamma^x_{xx}=\frac{\gamma^3X}{R^2}\!\left[-\frac{(1+v^2)M}{R}\!+\!\frac{(1-4\hspace*{1pt}v^2)\hspace*{1pt}M^2\!+\!v^2\,Q^2}{R^2}
\!+\!\frac{(M^2\!-\!Q^2)\,(y^2\!-\!X^2)}{R^4}\!+\!\frac{6\,v\,aMy}{R^3}\right]~,~~~~\label{xxx}    \\
\hspace*{-70pt}&&\Gamma^x_{yy}=\frac{\gamma X}{R^2}\left[\frac{M}{R}-\frac{M^2}{R^2}+\frac{(M^2-Q^2)\,(X^2-y^2)}{R^4}-\frac{6\,v\,aMy}{R^3}\right]~,~~~~\label{xyy} \\
\hspace*{-70pt}&&\Gamma^x_{tx}=\Gamma^x_{xt}=\frac{v\,\gamma^3X}{R^2}\left[\frac{(1\!+\!v^2\hspace*{0.9pt})\,M}{R}+\frac{3\,M^2-v^2\,Q^2}{R^2}
\!-\!\frac{(M^2\!-\!Q^2)\,(\hspace*{1pt}y^2\!-\!X^2\hspace*{1pt})}{R^4}\!-\!\frac{6\,v\,aMy}{R^3}\right]~,~~~~\label{xtx} \\
\hspace*{-70pt}&&\Gamma^x_{ty}=\Gamma^x_{xt}=\frac{v\hspace*{1pt}\gamma^2\hspace*{1pt}y}{R^2}\!\left[\frac{2M}{R}\!-\!\frac{M^2+Q^2}{R^2}
\!+\!\frac{2\hspace*{1pt}(M^2\!-\!Q^2)X^2}{R^4}\right]
\!-\!\frac{aM\gamma^2\left[R^2\!+\!3\hspace*{1pt}v^2\hspace*{1pt}(\hspace*{1pt}y^2\!-\!X^2\hspace*{1pt})\right]}{R^5} \hspace*{2pt}~,   \label{xty}  \\
\hspace*{-70pt}&&\Gamma^x_{xy}\!=\!\Gamma^x_{yx}\!=\!-\frac{\gamma^2\,y}{R^2}\!\left[\frac{(1\!+\!v^2)M}{R}\!-\!\frac{M^2\!+\!v^2\,Q^2}{R^2}
\!+\!\frac{2X^2(M^2\!-\!Q^2)}{R^4}\!-\!\frac{6\,v\,aMy}{R^3}\right]\!-\!\frac{2\,v\gamma^2aM}{R^3}~,     \label{xxy}  \\
\hspace*{-70pt}&&\Gamma^y_{tt}=\frac{\gamma^2\,y}{R^2}\left[\,\frac{(\hspace*{1pt}1\!+\!v^2\hspace*{1pt})\,M}{R}
-\frac{(4\!+\!v^2\hspace*{1pt})\,M^2\!+\!Q^2}{R^2}\!+\!\frac{v^2\,(M^2\!-\!Q^2)\,(\hspace*{1pt}y^2\!-\!X^2)}{R^4}\,\right]
-\frac{2\,v\,\gamma^2\,aM}{R^3}~\hspace*{2pt},  \label{ytt} \\
\hspace*{-70pt}&&\Gamma^y_{xx}=\frac{\gamma^2\,y}{R^2}\left[\,\frac{(1\!+\!v^2)\,M}{R}\!-\!\frac{(1\!+\!4\,v^2)\,M^2\!+\!v^2\,Q^2}{R^2}\!+\!\frac{(M^2\!-\!Q^2)\,(y^2\!-\!X^2)}{R^4}\,\right]
\!-\!\frac{2\,v\,\gamma^2\,aM}{R^3}~\hspace*{2pt},~~~~~~~~\label{yxx} \\
\hspace*{-70pt}&&\Gamma^y_{yy}=-\frac{M\,y}{R^3}+\frac{M^2\,y}{R^4}+\frac{(M^2\!-\!Q^2\,)\hspace*{4pt}(X^2-y^2\,)\hspace*{3pt}y}{R^6}~\hspace*{2pt},~~~~\label{yyy} \\
\hspace*{-70pt}&&\Gamma^y_{tx}=\Gamma^y_{xt}=\frac{2\,v\,\gamma^2\,y}{R^2}\left[-\frac{M}{R}+\frac{(M^2-Q^2)X^2}{R^4}+\frac{2M^2+Q^2}{R^2}\right]
+\frac{(1+v^2)\gamma^2aM}{R^3}~,  \label{ytx} \\
\hspace*{-70pt}&&\Gamma^y_{ty}=\Gamma^y_{yt}=\frac{v\,\gamma X}{R^2}\!\left[\frac{M}{R}-\frac{M^2}{R^2}+\frac{2\hspace*{1.5pt}(M^2-Q^2)\hspace*{1.5pt}y^2}{R^4}\right]~,~~~~\label{yty} \\
\hspace*{-70pt}&&\Gamma^y_{xy}=\Gamma^y_{yx}=\!-\frac{\gamma X}{R^2}\!\left[\frac{M}{R}-\frac{M^2}{R^2}+\frac{2\hspace*{1pt}(M^2\!-Q^2)\hspace*{1pt}y^2}{R^4}\right]~,~~~~\label{yxy}
\end{eqnarray}}
where we have neglected the third- and higher-order terms, the independence of the metric on $z$ has been taken into account, and the simplified form of
$R=\sqrt{\gamma^2(x-vt)^2+y^2-a^2}$ has been used. Notice that $R$ can be further approximated by $\sqrt{\gamma^2(x-vt)^2+y^2}$ for calculating the gravitational deflection up to second order.

\section{\emph{2PM} equations of motion derived by the Euler-Lagrange method} \label{B}

The explicit form of the \emph{Lagrangian} $L=-g_{\mu\nu}\dot{x}^\mu\dot{x}^\nu$ ($\mu,~\nu$ run over the values $0,~1,~2$)~\cite{WuckSperh2004} for a test particle propagating on the equatorial plane of the moving gravitational source is written as
{\small \begin{eqnarray}
\hspace*{-70pt}\nn L=\dot{t}^{\hspace*{1.3pt}2}\!\left[1\!-\!\frac{2(1\!+\!v^2)\gamma^2M}{R}\!+\!\frac{M^2\!+\!\gamma^2(M^2\!+Q^2)}{R^2}\!+\!\frac{4v\gamma^2aM X_2}{R^3}
-\frac{v^2\gamma^2(M^2\!-Q^2)\,X_1^2}{R^4}\right]\!-\dot{x}^2\!\times  \\
\hspace*{-70pt}\nn\hspace*{6pt}\left[1\!+\!\frac{2(1\!+\!v^2)\gamma^2M}{R}\!+\!\frac{M^2\!\!-\!v^2\gamma^2(M^2\!\!+\!Q^2)}{R^2}\!-\!\frac{4v\gamma^2aMX_2}{R^3}
\!+\!\frac{\gamma^2(M^2\!-\!Q^2)X_1^2}{R^4}\right]\!-\!\dot{y}^2\!\left[\!\left(\!1\!+\!\frac{M}{R}\right)^{\hspace*{-1.8pt}2} \right.  \\
\hspace*{-70pt}\nn\hspace*{6pt} \left. \!+\frac{(M^2\!-\!Q^2)\,X_2^2}{R^4}\right]
\!+2\,\dot{t}\,\dot{x}\left\{v\,\gamma^2\left[\frac{4\,M}{R}\!-\!\frac{M^2+Q^2}{R^2}\!+\!\frac{(M^2\!-\!Q^2)\,X_1^2}{R^4}\right]
-\frac{2(1\!+\!v^2)\gamma^2\,a\,M\,X_2}{R^3}\right\}  \\
\hspace*{-70pt}\hspace*{6pt}+\,2\,\dot{t}\,\dot{y}\left[\frac{2\,\gamma\,a M X_1}{R^3}\!+\!\frac{v\,\gamma\,(M^2\!-\!Q^2)X_1X_2}{R^4}\right]
\!-\!2\,\dot{x}\,\dot{y}\left[\frac{2\,v\gamma\,a M X_1}{R^3}\!+\!\frac{\gamma(M^2\!-\!Q^2)X_1X_2}{R^4}\right]~,  \label{L}
\end{eqnarray}}
where dots denote derivatives with respect to the trajectory parameter $\xi$ which has the same physical meaning as $p$. We substitute Eq.~(\ref{L}) into the Euler-Lagrange equation
\begin{eqnarray}
\frac{d}{d\xi}\frac{\partial L}{\partial\dot{q}}-\frac{\partial L}{\partial q}=0~, ~~~~~~(q=t,~x,~ y)~,
\end{eqnarray}
and get the equations of motion up to second post-Minkowskian order as follows:
{\small \begin{eqnarray}
\hspace*{-70pt}\nn 0=\ddot{t}+\frac{v\gamma^3\,\dot{t}^{\hspace*{1.3pt}2}X}{R^2}\!\left[-\frac{(1+v^2)M}{R}\!-\!\frac{v^2(7\!+\!v^2)\,\gamma^2M^2\!-\!Q^2}{R^2}
\!+\!\frac{v^2(M^2\!-\!Q^2)\,(y^2\!-\!X^2)}{R^4}\!+\!\frac{6\,v\,aMy}{R^3}\right]  \\
\hspace*{-70pt}\nn\!+\frac{\gamma^3\dot{x}^2X}{R^2}\!\left[\!\frac{v(v^2\!-\!3)M}{R}\!-\!\frac{v[(3\!+\!9v^2\!-\!4v^4)\gamma^2M^2\!+\!(v^2\!-\!2)Q^2]}{R^2}
\!+\!\frac{v(M^2\!-\!Q^2)(y^2\!-\!\!X^2)}{R^4}\!+\!\frac{6aMy}{R^3}\!\right]  \\
\hspace*{-70pt}\nn+\,\frac{2\,\gamma^3\,\dot{t}\,\dot{x}\,X}{R^2}\!\left[\frac{(1+v^2)\,M}{R}+\frac{v^2\,(7+v^2)\,\gamma^2\,M^2-Q^2}{R^2}
-\frac{v^2\,(M^2-Q^2)\,(y^2\!-\!X^2)}{R^4}-\frac{6\,v\,aMy}{R^3}\right]  \\
\hspace*{-70pt}+\frac{2\,(1+v^2)\,\gamma^2\,M\,\dot{t}\,\dot{y}\,y}{R^3}-\frac{4\,v\gamma^2\,M\,\dot{x}\,\dot{y}\,y}{R^3}+\frac{4\,v\gamma^2M\,\ddot{x}}{R}~, \label{L1}
\end{eqnarray}
\begin{eqnarray}
\hspace*{-70pt}\nn 0=\ddot{x}+\frac{\gamma^3\,\dot{t}^{\hspace*{1.3pt}2}\,X}{R^2}\left[\frac{(\,1-3\,v^2\,)\,M}{R}
+\frac{(-\,4+9\,v^2+3\,v^4\,)\,\gamma^2\,M^2+(\,2\,v^2-1\,)\,Q^2}{R^2}+\frac{6\,v^3\,aMy}{R^3} \right.\\
\hspace*{-70pt}\nn\left.+\frac{v^2(M^2\!-\!Q^2)\,(y^2\!-\!X^2)}{R^4}\right]
\!+\!\frac{\gamma^3\,\dot{x}^2\,X}{R^2}\!\left[-\frac{(1+v^2)\,M}{R}+\frac{(1+7\,v^2)\,\gamma^2\,M^2+v^2Q^2}{R^2}+\frac{6\,v\,aMy}{R^3}  \right. \\
\hspace*{-70pt}\nn\left.+\,\frac{(M^2\!-\!Q^2)\,(y^2\!-\!X^2)}{R^4}\right]\!+\!\frac{2\,v\,\gamma^3\,\dot{t}\,\dot{x}\,X}{R^2}
\left[\,\frac{(1+v^2)\,M}{R}\!-\!\frac{(1+7\,v^2)\,\gamma^2\,M^2\!+v^2\,Q^2}{R^2}-\frac{6\,v\,aMy}{R^3}  \right. \\
\hspace*{-70pt}\left.-\,\frac{(M^2-Q^2)\,(y^2-X^2)}{R^4}\right]+\frac{2\,\gamma\,M\,(v\,\dot{X}_0-\dot{X})\,\dot{y}\,y}{R^3}-\frac{4\,v\,\gamma^2\,M\,\ddot{t}}{R}~, \label{L2}
\end{eqnarray}
\begin{eqnarray}
\hspace*{-70pt}\nn0=\ddot{y}+\dot{t}^{\hspace*{1.3pt}2}\left\{\frac{\gamma^2\,y}{R^2}\left[\frac{(1+v^2)M}{R}\!-\!\frac{(4+v^2)M^2\!+Q^2}{R^2}
\!+\!\frac{v^2(M^2\!-Q^2)\,(y^2-\!X^2)}{R^4}\right]\!-\!\frac{2\,v\,\gamma^2\,aM}{R^3}\right\}  \\
\hspace*{-70pt}\nn+\,\dot{x}^2\!\left\{\frac{\gamma^2\,y}{R^2}\left[\,\frac{(\,1+v^2\,)\,M}{R}\!-\!\frac{(\,1+4\,v^2\,)\,M^2+v^2\,Q^2}{R^2}\!+\!\frac{(M^2-Q^2)\,(y^2-X^2)}{R^4}\,\right]
\!-\!\frac{2\,v\,\gamma^2\,aM}{R^3}\right\}  \\
\hspace*{-70pt}+\,2\gamma^2\,\dot{t}\,\dot{x}\!\left[-\frac{2vMy}{R^3}\!+\!\frac{v(5M^2\!+Q^2)y}{R^4}
\!-\!\frac{v(M^2\!-\!Q^2)(y^2\!-\!X^2)y}{R^6}\!+\!\frac{(1\!+\!v^2)aM}{R^3}\right]\!-\!\frac{2M\dot{X}\dot{y}X}{R^3}~, \hspace*{30pt} \label{L3}
\end{eqnarray}}
where $R$ can be approximated by $\sqrt{X^2+Y^2}$ for the deflection up to second order. Here we have also regarded the order of $\dot{y}$ and $\ddot{y}$ as $\Phi$ and ignored the third- and higher-order terms. The main relations used in the calculations of Eqs.~(\ref{L1}) - (\ref{L3}) are given as follows:
\begin{eqnarray}
\dot{X}_0=\dot{T}=\gamma(\dot{t}-v\dot{x})~,~~~~\dot{X}_1=\dot{X}=\gamma(\dot{x}-v\dot{t})~,~~~~\dot{X}_2=\dot{Y}=\dot{y}~,  \label{relation-1}  \\
\frac{\partial R}{\partial t}=-\frac{v\gamma X}{R}~,  \label{relation-2}  \\
\frac{\partial}{\partial t}\frac{X^2}{R^4}=\frac{2v\gamma X(X^2-y^2)}{R^6}~, \label{relation-3}  \\
\frac{\partial R}{\partial x}=\frac{\gamma X}{R}~, \label{relation-4}  \\
\frac{\partial}{\partial x}\frac{X^2}{R^4}=\frac{2\gamma X(y^2-X^2)}{R^6}~, \label{relationship-5}  \\
\frac{\partial R}{\partial y}=\frac{y}{R}~, \label{relation-6}  \\
\frac{\partial}{\partial y}\frac{y}{R^3}=\frac{X^2-2y^2}{R^5}~, \label{relation-7}  \\
\frac{dR}{d\xi}=\frac{X\dot{X}+y\dot{y}}{R}~, \label{relation-8}  \\
\frac{d}{d \xi}\frac{X}{R^3}=\frac{R^2\dot{X}-3X(X\dot{X}+y\dot{y})}{R^5}~, \label{relation-9}  \\
\frac{d}{d \xi}\frac{y}{R^3}=\frac{R^2\dot{y}-3y(X\dot{X}+y\dot{y})}{R^5}~, \label{relation-10}  \\
\frac{d}{d \xi}\frac{X^2}{R^4}=\frac{2R^2X\dot{X}-4X^2(X\dot{X}+y\dot{y})}{R^6}~, \label{relation-11} \\
\frac{d}{d \xi}\frac{y^2}{R^4}=\frac{2R^2y\dot{y}-4y^2(X\dot{X}+y\dot{y})}{R^6}~, \label{relation-12}  \\
\frac{d}{d \xi}\frac{Xy}{R^4}=\frac{R^2(\dot{X}y+X\dot{y})-4Xy(X\dot{X}+y\dot{y})}{R^6}~. \label{relation-13}
\end{eqnarray}

In order to compare Eqs.~(\ref{L1}) - (\ref{L3}) with Eqs.~(\ref{geodesic-t}) - (\ref{geodesic-y}), we need to calculate the explicit forms for $\ddot{t}$ and $\ddot{x}$ up to \emph{1PM} order, which are not difficult to be obtained by solving Eqs.~(\ref{L1}) - (\ref{L2}) in the \emph{1PM} approximation as
\begin{eqnarray}
\ddot{t}=\frac{\gamma^3\left[v(1+v^2)\dot{t}^{\hspace*{1.3pt} 2}-2(1+v^2)\dot{t}\dot{x}+v(3-v^2)\dot{x}^2\right] M X}{R^3}~,    \label{first-order-t} \\
\ddot{x}=\frac{\gamma^3\left[(1+v^2)\dot{x}^2-2v(1+v^2)\dot{t}\dot{x}-(1-3v^2)\dot{t}^{\hspace*{1.3pt} 2}\right] M X}{R^3}~.   \label{first-order-x}
\end{eqnarray}
Substituting Eqs.~(\ref{first-order-t}) and (\ref{first-order-x}) into Eqs.~(\ref{L2}) and (\ref{L1}), respectively, one finds that Eqs.~(\ref{L1}) - (\ref{L3}) take the same form as Eqs.~(\ref{geodesic-t}) - (\ref{geodesic-y}) derived via calculating the Christoffel symbols.

\end{document}